\documentclass{article}
\usepackage{amssymb}

\newcommand{\fr}{\frac}

\newcommand{\be}{\begin{equation}}
\newcommand{\ee}{\end{equation}}
\newcommand{\ba}{\begin{array}}
\newcommand{\ea}{\end{array}}
\newcommand{\beqa}{\begin{eqnarray}}
\newcommand{\beqaa}{\begin{eqnarray*}}
\newcommand{\al}{\alpha}
\newcommand{\ald}{{\dot{\alpha}}}
\newcommand{\bet}{\beta}
\newcommand{\betd}{{\dot{\beta}}}

\newcommand{\gad}{{\dot{\gamma}}}
\newcommand{\ga}{\gamma}
\newcommand{\la}{\lambda}
\newcommand{\La}{\Lambda}
\newcommand{\lab}{\bar{\lambda}}
\newcommand{\Lab}{\bar{\Lambda}}
\newcommand{\si}{\sigma}

\newcommand{\sib}{{\bar{\sigma}}}
\newcommand{\xib}{{\bar{\xi}}}
\newcommand{\zeb}{{\bar{\zeta}}}
\newcommand{\te}{\theta}
\newcommand{\teb}{{\bar{\theta}}}

\newcommand{\del}{\partial}
\newcommand{\eeqa}{\end{eqnarray}}
\newcommand{\eeqaa}{\end{eqnarray*}}
\newcommand{\ep}{\epsilon}

\newcommand{\Dc}{{\cal D}}
\newcommand{\Kc}{{\cal K}}
\newcommand{\Nc}{{\cal N}}

\newcommand{\Oc}{{\cal O}}

\newcommand{\Omo}{\Omega}
\newcommand{\omo}{\omega}

\newcommand{\omob}{{\bar{\omega}}}
\newcommand{\kd}{\delta}

\newcommand{\rD}{{\rm D}}
\newcommand{\ved}{\wedge}

\begin{document}
\begin{center}

{\Huge Superfield formulation of N=4 supersymmetric Yang--Mills theory in extended superspace}

\vspace{17mm}

\"{O}mer F. Dayi$^{a,b,}$\footnote{E-mail addresses: dayi@itu.edu.tr and
dayi@gursey.gov.tr.} \,{and}\, Kayhan \"{U}lker
$^{b,}$\footnote{E-mail address: kulker@gursey.gov.tr.}\\
\vspace{5mm}

\end{center}

\noindent {\em $^{a}${\it Physics Department, Faculty of Science and
Letters, Istanbul Technical University, TR-34469 Maslak--Istanbul, Turkey.} }

\vspace{3mm}

\noindent {\em $^{b}${\it Feza G\"{u}rsey Institute, P.O. Box 6,
TR--34684, \c{C}engelk\"{o}y--Istanbul, Turkey. } }

\vspace{2cm}

{\small
\noindent
Action of 4 dimensional N=4 supersymmetric Yang--Mills theory 
is written by employing the 
superfields in N=4 superspace which were used to prove the equivalence of its constraint equations
and  equations of motion. Integral forms
of the extended superspace are engaged to collect all of the superfields in
one ``master" superfield.  The proposed N=4  supersymmetric Yang--Mills action in 
extended superspace is shown to acquire a simple form 
in terms of the master superfield.}

\newpage


\section{Introduction}

Maximally supersymmetric gauge model in four dimensions that contains fields with spins at most one
is N=4 supersymmetric Yang--Mills (SYM)  theory \cite{gso}.
This theory is distinguished for its finiteness and duality properties and studied 
extensively since last three decades (for some reviews see \cite{soh,dvec}). In spite of 
these 
facts, a superfield formulation of N=4 SYM in extended superspace is still lacking. An  
off-shell formulation in terms of  auxiliary fields of the N=4 SYM multiplet is still unknown 
(for a formulation with an infinite number
of fields see \cite{ssw}) A progress made in this direction was to accomplish 
the equivalences of the superfield constraint equations and the
equations of motion for  N=3 and N=4 SYM theories \cite{soh78,wit78}. 
In \cite{hhls} a complete proof of this equivalence relation  for N=3 SYM was given by introducing a suitable gauge choice  which eliminates  gauge freedom depending on  Grassmann coordinates.
Obviously, by  superfields we mean fields written as functions
of superspace variables. Indeed,  this gauge choice permits one to find some recursion relations from the constraint 
equations to construct superfields order by order in Grassmann variables. This method  was 
also applied to ten dimensional SYM equations \cite{hs}.  Unlike the accustomed superfields, 
components of the superfields constructed in \cite{hhls,hs} 
do not encompass any auxiliary field. Hence, they do not demand an off--shell supersymmetric 
formulation, but at the cost of considering superfields 
which do not possess the usual supersymmerty properties.

Superfields constructed in \cite{hs} were employed to construct physical states
in Berkovits  quantization of superparticles and superstrings in ten dimensions \cite{ber}. 
Also, the approach of \cite{hhls} was
applied to define  deformed  N=4 SYM equations \cite{sw}.
Recently,  in terms of these superfields
an alternative superfield formulation  of N=1 
SYM without auxiliary fields in 4 dimensions was given \cite{dayi}. 

We present a formulation of 4 dimensional N=4 SYM
action in terms of the superfields of \cite{hhls}. Moreover, we show that
these fields can be written as integral forms and be collected in
a ``master" superfield such that the N=4 action can be expressed in a simple, compact form.

Though how to determine components of the superfields by recursion relations
is known, actual
calculation of components
which are third or fourth order in Grassmann variables is a hard  task. 
One of the other  important issues to write an action in extended superspace is to define a measure which is invariant under 
the global $SU(4)$ group. 
We propose a measure which is suitable to achieve our goal.
Acquainted with these we write an action in terms of superfields
and prove that the action of N=4 SYM theory in terms of component fields results,
after a lengthy calculation.
For the coefficients appearing in the action  there are
more than one solution. 

Engaging  differentials  of  Grassmann variables 
  the N=4 SYM
superfields can be written as integral forms \cite{m} (see also \cite{zp}) and be collected into a ``master" superfield. Then, the N=4 SYM action can be written in a compact way.  This action is 
apparently first order in space--time derivatives and there are two terms which are
quadratic and cubic in master superfield. Indeed, all other powers of  master superfield
give vanishing contributions
to the action. Also in this case,
there are some different solutions for the coefficients involved.

In the next section we recall  formulation of 4 dimensional
N=4 SYM  theory in  terms of component fields.
In Section 3 after giving the definitions of superfields, we present 
their first two components in  Grassmann variables of the extended superspace.
The higher components are listed in  Appendix.
In Section 4 we present the general formulation in terms of superfields
after a choice of measure.  
In Section 5 
we give the definition of master superfield as a collection of 
integral forms. We
demonstrate  that  
the N=4 SYM theory action in  extended superspace acquires a simple form. In the last section we discuss the
results obtained and some open questions.

\section{Component fields formulation}

The N=4 Yang--Mills supermultiplet consists of one gauge field\footnote{We always make the identification 
$x_{\al\ald}=\si_{\al\ald}^\mu x_\mu \, ,\,
x^{\ald \al}= \sib^{\ald\al}_{\mu} x^{\mu}\, ,\, \del_{\al\ald}=\si_{\al\ald}^\mu \del_\mu$ . } 
$a_{\al\ald} = \sigma_{\al\ald}^{\mu} A_\mu $, 
eight Weyl fermions $\la_{i\al}  \, ,\, \lab^i_\ald$ 
and six scalars $\phi_{ij}=-\phi_{ji}=\fr{1}{2}\ep_{ijkl}\phi^{kl}$. 
Spinor indices are\footnote{We use Wess-Bagger conventions \cite{wb} to raise and lower the spinor indices:
$\te_\al =\ep_{\al\bet}\te^\bet \, ,\, \te^\al =\ep^{\al\bet}\te_\bet \, ,\,  \ep_{\al\bet}\ep^{\bet\gamma}=\delta_{\al}^{\gamma}$.}
$\al\, ,\, \ald =1,2.$ 
$i,j=1,\cdots,4,$ denote indices of the global symmetry group $SU(4) .$ 
In fact, $a_{\al\ald}$ is a singlet,  
$\la_i^\al$ and $\lab^i_\ald$ are in the  $\mathbf{{4}}$ and 
$\mathbf{\bar{4}}$ representation and
$\phi_{ij}$ are in the second rank, self dual $\mathbf{6}$ representation of $SU(4).$
All of the fields are in the adjoint representation of a non--abelian gauge group.

Hermitian conjugation which we attribute to the fields is 
$$
(a_{\al\betd})^\dag = - a_{\bet\ald}\, ,\, 
(\la_{i\al})^\dag = \lab^i_\ald \, ,\, (\phi_{ij})^\dag =\phi^{ij}.
$$

N=4 extended SYM action in the component fields $a_{\al\ald},\la_{i\al},\phi_{ij}$ can be written as 
\beqa
I & =&  \int\  d^4x {\rm Tr}{\big (}
\frac{1}{8} f_{\ald\betd}f^{\ald 
\betd}  +
\frac{1}{8} f_{\al\bet}f^{\al \bet} + 
\frac{1}{16}{\rm D}_{\al\ald}\phi_{ij}{\rm D}^{\ald\al}\phi^{ij} 
-\fr{i}{4}\la_i^\al {\rm D}_{\al\ald}\lab^{i\ald} \nonumber \\
& & 
-\fr{i}{8}\phi^{ij}\{\la_i^\al , \la_{j\al\}}  
-\fr{i}{8}\phi_{ij}\{\lab^i_\ald , \lab^{j\ald} \}    
+\frac{1}{64}[\phi_{ij},\phi_{kl}] 
[\phi^{ij},\phi^{kl}]
{\big )} ,\label{n4c}
\eeqa
where $D_{\al\ald}=\del_{\al\ald} + [ a_{\al\ald}, \cdot ].$ $ f_{\al\bet}$ and $f_{\ald\betd}$ 
are self-dual and anti-self-dual field strengths defined as\footnote{Note that, 
$f_{\al\bet}= (\sigma^{\mu\nu})_\al^\gamma \epsilon_{\gamma\bet}f_{\mu\nu}$ and 
$f_{\ald\betd} = \epsilon_{\ald\gad} (\bar{\sigma}^{\mu\nu})^\gad_\betd f_{\mu\nu}$ where 
$f_{\mu\nu} = \del_\mu a_\nu -\del_\nu a_\mu + [ a_\mu , a_\nu ] $ as usual.} 
$$f_{\al\bet}=-\fr{1}{2} \ep^{\ald\betd}\left( \del_{\al\ald}a_{\bet \betd}
-\del_{\bet \betd}a_{\al\ald}
+[ a_{\al\ald}, a_{\bet \betd}]\right) 
 ,$$
$$f_{\ald\betd}=-\fr{1}{2} \ep^{\al\bet}\left( \del_{\al\ald}a_{\bet \betd}
-\del_{\bet \betd}a_{\al\ald}
+[ a_{\al\ald}, a_{\bet \betd}] 
\right).$$

The action (\ref{n4c}) is invariant under the on-shell N=4 supersymmetry transformations:
\beqa
\kd a_{\al \ald} & = & -\xi_{i\al}\lab^i_\ald -\xib _\ald^i\la_{i\al}, \label{ss1} \\
\kd\la_{i\al} & = & 2i \xi_i^\bet f_{\al\bet}-i\xi_{j\al}
[\phi_{ik},\phi^{kj}]-2i\xib^{j\ald}D_{\al\ald}\phi_{ij},\\
\kd\phi_{ij} & = & \xi_i^\al \la_{j\al} -\xi_j^\al\la_{i\al}
+\ep_{ijkl} \xib^k_\ald \lab^{l\ald}, \label{ss3}
\eeqa
where $\xi_i,\xib^i$ are constant Weyl spinors.

\section{Superfields in N=4 superspace}

N=4 extended superspace is parametrized by the coordinates
\be
\label{n4sp}
(x^\mu, \te_i^\al, \teb_\ald^i ).
\ee

Translations in this extended  superspace
$$
x_{\al\ald}\rightarrow x_{\al\ald}  +2i( \zeb^{i}_{\ald}\te_{i\al} + \zeta_{i\al}\teb_\ald^i)
\quad ,\quad
\te_i^\al \rightarrow  \te_i^\al + \zeta_i^\al
\quad ,\quad
\teb_\ald^i \rightarrow \teb_\ald^i +\zeb^{i}_{\ald} 
$$
are generated by $T\equiv \zeta_{i}^{\al}Q_\al^i +\zeb^{i}_{\ald}{\bar{Q}_{ i}}^{\ald}$,
where $\zeta_{i}^{\al},\zeb^{i\ald}$ are Grassmann constants. The  supercharges 
\be
\label{charges}
Q_\al^i  =  \fr{\del}{\del\te_i^\al}- i\teb^{i\ald}\del_{\al\ald} 
\quad ,\quad 
\bar{Q}_{i\ald}  =  -\fr{\del}{\del\teb^{i\ald}}+ i\te_i^\al\del_{\al\ald},
\ee
satisfy the graded algebra  
\beqaa
\{Q_\al^i,\bar{Q}_{\ald j}\}&  =& 2i \delta^i_j \del_{\al\ald} ,\\
\{Q_\al^i,Q_\bet^j\}  = 
\{\bar{Q}_{\ald i},\bar{Q}_{\betd j}\} &=& {[\del_{\al\ald} ,Q_\bet^i]} = {[ \del_{\al\ald},\bar{Q}_{\betd i} ]} = 0.
\eeqaa
To construct supersymmetric actions in superspace 
it is convenient to be acquainted with
the differential operators 
\be
{\rm D}_\al^i  =  \fr{\del}{\del\te_i^\al}+i\teb^{i\ald}
\del_{\al\ald}\quad ,\quad
\bar{\rm D}_{i\ald}  =  -\fr{\del}{\del\teb^{i\ald}}-i\te_i^\al 
\del_{\al\ald},
\ee
that anticommute with the supercharges (\ref{charges}):
$$ \{Q_\al^i,{\rm D}_\bet^j\}  = 
\{\bar{Q}_{\ald i},{\rm D}_\al^j\}=
\{Q_\al^i,\bar{\rm D}_{j\ald}\}  = 
\{\bar{Q}_{\ald i},\bar{\rm D}_{\betd j}\}=0.
$$

An off--shell  N=4  SYM formulation is not available which would lead 
to construction of N=4 superfields living in the N=4 superspace (\ref{n4sp})
making use of accustomed methods. 
However, there exists another approach of introducing superfields whose components are constituted by the fields which are not auxiliary, in  terms of the constraint equations for the 
superconnections $A_{\al\ald},$ $\omo^i_\al $ and $\omob_{i\ald} $ \cite{hhls}.  The
supercovariant derivatives in N=4 superspace\footnote{Note that, $(A_{\al\betd})^\dag = -A_{\bet\ald}$ but $(\omo^i_\al)^\dag =\omob_{i\ald}$},
\beqa
\nabla^i_\al & = & {\rm D}_\al^i+[\omo^i_\al ,\cdot ] ,\\
\bar{\nabla}_{i\ald} & = & \bar{\rm D}_{i\ald}-[\omob_{i\ald} ,\cdot ]  ,\\
\nabla_{\al\ald} & = & \del_{\al\ald} +[A_{\al\ald},\cdot].
\eeqa
should satisfy
the constraint equations
\beqa
\{ \nabla_\al^i ,\bar{\nabla}_{j\ald} \} &=&-2i\kd^i_j \nabla_{\al\ald} , \label{c1} \\
\{\nabla_\al^i,\nabla_\bet^j\}=-2i\ep_{\al\bet}\Phi^{ij} &,& 
\{\bar{\nabla}^\ald_i,\bar{\nabla}^\betd_j\}=2i\ep^{\ald\betd}\Phi_{ij} , \\
{ [\nabla_\al^i , \nabla_{\bet\betd} ]}=\ep_{\al\bet}\Lab_\betd^i &,& 
{[\bar{\nabla}_{i\ald},\nabla_{\bet\betd} ]}=-\ep_{\ald\betd}\La_{i\bet}.  \label{c3}
\eeqa
Here the upper--case letters indicate superfields whose first components are
proportional to the component fields given by the lower--case letters.
Let us also define the operator
\be
\label{Dc1}
\Dc =\te_i^\al \nabla_\al^i
-\teb^{i\ald}\bar{\nabla}_{i\ald}.
\ee
One can show that Bianchi identities 
resulting from (\ref{c1})--(\ref{c3}) lead to the recursion relations
\beqa
\Dc A_{\al\ald}&=&
-\te_{i\al}\Lab_\ald^i
-\teb^i_\ald\La_{i\al} , \label{reca}\\
\Dc \La_{i\al} & = & 
2i\te_i^\bet F_{\al\bet}
-i[\Phi_{ik},\Phi^{kj} ]\te_{j\al}
-2i\teb^{j\ald}\nabla_{\al\ald}\Phi_{ij}, \label{recl}\\
\Dc \Phi_{ij}  & = & 
\te_i^\al\La_{j\al}
-\te_j^\al\La_{i\al} +\ep_{ijkl} \teb^k_\ald\Lab^{l\ald}. \label{recp}
\eeqa

Being superconnections there are some redundant parts in
$\omo,\omob$ which should be eliminated, obviously leaving the usual Yang--Mills 
gauge transformations  intact. Adopting the gauge fixing condition 
\be\label{gauge}
\te_i^\al\omo^i_\al + \teb^{i\ald} \omob_{i\ald} =0,
\ee
that eliminates all the gauge transformations depending on the Grassmann coordinates
$\te_i^\al ,\ \teb^{i\ald},$
which is similar to the Wess--Zumino condition,
the recursion relations for the spinor superconnections 
can be derived from (\ref{reca}) and (\ref{recp}) as
\beqa
(1+\Dc)\omo^i_\al & = & 2i\teb^{i\ald}A_{\al\ald} -2i\Phi^{ij}\te_j^\al , \label{reco}\\ 
(1+\Dc)\omob_{i\ald} & = & 2i\te_i^\al A_{\al\ald}+2i\Phi_{ij} \teb^j_\ald .\label{recob}
\eeqa

Note that after the gauge choice (\ref{gauge}) the operator (\ref{Dc1}) turned to be the counting operator of the anticommuting coordinates $\te_i^\al $ and $\teb^i_\ald :$
\[
\Dc =\te_i^\al \fr{\del}{\del \te^\al_i}
+\teb^{i\ald} \fr{\del}{\del \teb^{i\ald}}.
\]
Therefore, the superfields $\omo,A,\Phi ,\La$ 
can be found from the recursion relations (\ref{reca})--(\ref{recob}) order by order in $\te ,\ \teb . $

When one  replaces the upper--case letters with the lower--case ones
in (\ref{reca})--(\ref{recl}), $\Dc $ can be replaced with $\delta$ which is the supersymmetry transformation
(\ref{ss1})--(\ref{ss3}) with the replacements $\xi \rightarrow \te ,$ $\xib \rightarrow \teb .$
If the above superfields are written order by order in $\te,\teb ,$ as
\beqa
A_{\al\ald}&=&s_0 A^{(0)}_{\al\ald}+s_1 A^{(1)}_{\al\ald} + \cdots + s_n A^{(n)}_{\al\ald},\\
\Phi_{ij}&=&e_0 \Phi^{(0)}_{ij} +e_1 \Phi^{(1)}_{ij} +\cdots + e_n \Phi^{(n)}_{ij},\\
\La_{i\al}&=&z_0 \La^{(0)}_{i\al} +z_1 \La^{(1)}_{i\al} \cdots + z_n \La^{(n)}_{i\al},
\eeqa
where $s_m , e_m , z_m \, ; m=0,1 \cdots 16,$ are real constants, the unique solution to any desired order  can also be found as,
\be
\label{dt}
A^{(m)}_{\al\ald}=\delta A^{(m-1)}_{\al\ald} \quad , \quad 
\Phi^{(m)}_{ij} = \delta \Phi^{(m-1)}_{ij} \quad , \quad
\La^{(m)}_{i\al} =\delta \La^{(m-1)}_{i\al}
\ee
by setting $s_0=e_0=z_0=1$ and
\be
s_m = e_m \quad , \quad   mz_m =s_{m-1}   \quad , \quad ms_m =z_{m-1};
\quad  \quad m=1,\cdots, 16 .
\ee

Hence, to obtain
the superfields $A,\La ,\Phi$ one can proceed in two
equivalent ways: Make use of the recursion relations (\ref{reca})--(\ref{recl}) directly or  
perform the transformations (\ref{dt}).

In terms of the arbitrary scale factors $l$ and  $b$ which are real constants,
let us define the zeroth order components as 
$$A^{(0)}_{\al\ald}=a_{\al\ald},\  \La^{(0)}_{i\al}=l\la_{i\al} ,\ \Phi^{(0)}_{ij}=b\phi_{ij} .$$
The first order components of the superfields $A,\Phi ,\La$ can be derived from these as
\beqa
A^{(1)}_{\al\ald}&=&-l\te_{i\al}\lab_\ald^i 
-l\teb^i_\ald\la_{i\al} ,\label{a1} \\
\La^{(1)}_{i\al} & = & 2i\te_i^\bet f_{\al\bet}
-ib^2[\phi_{ik},\phi^{kj} ]\te_{j\al}
-2ib\teb^{j\ald} D_{\al\ald}\phi_{ij}, \label{l1}\\
\Phi^{(1)}_{ij}&=& l\te_i^\al\la_{j\al}
-l\te_j^\al\la_{i\al} + l\ep_{ijkl} \teb^k_\ald\lab^{l\ald}. \label{p1}
\eeqa

On the other hand, the spinor superconnection  $\omo $
can be separated 
into two parts:
$$\omo^i_\al=v^i_\al +u^i_\al ,$$
such that the gauge condition (\ref{gauge}) takes the form
$$
\te_i^\al v^i_\al+\teb^{i\ald} \bar{v}_{i\ald} =0\quad ,\quad \te_i^\al u^i_\al=\teb^{i\ald} 
\bar{u}_{i\ald} =0.
$$ 
There are no zeroth order components,
their  first and the second order components can be calculated 
from the recursion relations (\ref{reco})--(\ref{recob}) and 
(\ref{a1})--(\ref{p1})
as
\beqa
v^{(1)i}_{\al}  =  i\teb^{i\ald} a_{\al \ald} &,& 
v^{(2)i}_{\al}  =  \fr{-2il}{3}\teb^{i\ald}(\te_{k\al}\lab^k_\ald +\teb^k_\ald\la_{k\al} ), 
\label{o1}\\
u^{(1)i}_{\al}  = - ib\te_{j\al}\phi^{ij} &,&
 u^{(2)i}_{\al} = \fr{2il}{3}\te_{j\al}(\teb^{i\ald}\lab^j_\ald-\teb^{j\ald}\lab^i_\ald ).
\label{o2}
\eeqa

Here we presented the first two components of the superfields. The higher order terms are
listed  in Appendix.

\section{N=4 SYM action in terms of  superfields}

We wish to find an action in terms of the superfields $\omo,A,\La,\Phi$ and the derivatives
$\del_{\al\ald} ,$ such that after performing integrals over the Grassmann variables
$\te,\teb$ it attains the action in terms of component fields (\ref{n4c}).
Inspecting  components of the superfields $\omo,A,\La,\Phi $ one observes that if we do not
restrict the integration over $\te ,\teb$ but integrate 
over the whole superspace, the desired result cannot be achieved.

We  propose the action, in terms of the constant parameters $ k_1,\cdots ,k_6, $
\beqa
S & = & ik_1<\bar{\omo}_{i\ald}\del^{\ald \al} \omo^{i}_{\al}> 
 +ik_2<\bar{\omo}_{i\ald} [A^{\ald \al},\omo^i_\al] >  \nonumber \\
&&+ ik_3<\omo^{i\al}\La_{i\al}-\bar{\omo}_{i\ald}\Lab^{\ald i} > 
+k_4<A_{\al\ald}A^{\ald \al}> \nonumber \\
&&
+ ik_5<\Phi_{ij} \{\omo^{i\al} ,\omo_\al^j\}+\Phi^{ij} \{\omob_{i\ald}, \omob_j^\ald\}>
 +k_6<\Phi_{ij} \Phi^{ij}>  , \label{acg}
\eeqa
where we defined, by the normalization constant $\Nc = 1/3200,$ 
\be
\label{mes0}
<\Oc > \equiv \Nc \left( \int\ d^4x\ 
d\te_i^\al d\te_{j\al} d\teb^i_\ald d\teb^{j\ald}\ {\rm Tr}\ \Oc \right)_{\te=\teb=0}.
\ee
Thus, the only non--vanishing  $\te,\teb$ contribution to the integral is
\be
\label{mes}
<\te_i^\al \te_j^\bet \teb^{k\ald} \teb^{l\betd} \Kc (x)>
=\fr{\Nc}{8} \ep^{\al\bet}\ep^{\ald\betd}
(\kd^k_i\kd^l_j +\kd^k_j\kd^l_i)\int d^4x\ {\rm Tr}\ \Kc (x) ,
\ee
for any function $\Kc (x).$ 
With this choice of measure (\ref{mes0}), due to mass dimensions and R-charges of the superfields (Table 1),
(\ref{acg}) is the most general action one can write up to total derivatives.

\begin{table}[hbt]
\caption[t1]{ Dimensions   $d$,   and R-weights.}
\centering
\begin{tabular}{|c||c|c|c|c|c|}
\hline
&$A_{\al\ald}$&$\la_i$&$\Phi_{ij}$&$\omo^i $&$\te_i$ \\
\hline\hline
$d$&1&3/2&1&1/2&-1/2 \\
\hline
$R$&0&-1&-2&1&-1 \\
\hline
\end{tabular}
\end{table}

Because of  the choice of measure (\ref{mes0}) which is manifestly $SU(4)$ invariant, 
components of the superfields at most up to the fourth order
in $\te,\teb$ are required.  Carrying out integrals over the variables $\te,\teb$ 
in (\ref{acg})  is a very lengthy calculation although it is straightforward. 
Nevertheless, using the identity
\be
[\phi_{ij},\phi^{jk}]
[\phi_{kl},\phi^{li}]=\fr{1}{2}
[\phi^{ij},\phi^{kl}]
[\phi_{ij},\phi_{kl}] ,
\ee
and performing the integrals over $\te,\teb,$ we conclude that 
to get the action (\ref{n4c}) from (\ref{acg}) the coefficients 
$ k_1,\cdots ,k_6, $
should satisfy
the equations
\beqa
12k_2-3k_1 -4k_3+2k_4& = & 0 ,\\
-3k_1+10k_3-8k_4& = & \fr{3}{20\Nc } ,\\
k_2-2k_5 & = & 0 ,\\
k_4-2k_6& = & -\fr{1}{10\Nc l^2 },\\
3k_1-10k_3 +16k_6& = & \fr{3}{10\Nc b^2 }, \\
-k_2 +k_3 -k_4 +2k_5 +2k_6 & = & -\fr{3}{20\Nc bl^2 } ,\\
-3k_2 -7k_3 -3k_4 +18k_5 +16k_6 & = & \fr{3}{16\Nc b^4} .
\eeqa
Although these equations possess some different solutions, by fixing
\be
\label{cso}
k_6 = -\frac{3k_4}{2} 
\ee
one obtains the solution
\beqa
k_1 &=& -8 (104+b (282+b (16+63 b))),\\
k_2 &=& \frac{1}{10} (-2815-4 b (1974+b (115+441 b))),\\
k_3 &=&-\frac{12}{5} (105+b (282+b (20+63 b))), \\
k_4 &=&-3 \left(21+4 b^2\right),\\
k_5 &=& \frac{k_2}{2} ,
\eeqa
with the scale factors 
\be\label{scale}
b=\sqrt{-2+\sqrt{26}/2}\quad , \quad l=4 \sqrt{(5-4 b^2)/39} ,
\ee
whose signs can be taken diversely, i.e. $b\rightarrow \pm b,\ l\rightarrow \pm l$ are
also solutions.

\section{A formalism by integral forms }

To acquire an understanding of  underlying geometrical aspects of the formulation
given in the previous section, we would like to write superfields as  
integral forms \cite{m,zp}. Let us introduce
the differentials $d\te ,d\teb$ whose (wedge) products are commutative\footnote{Here, we write the wedge product symbol $\ved$ explicitly to avoid the notational confusion.}:
\beqaa
d\te_i^\al \ved d\te_j^\bet & = & d\te_j^\bet \ved d\te_i^\al , \\
d\teb^{i\ald} \ved d\teb^{j\betd} & = &d\teb^{j\betd} \ved d\teb^{i\ald}, \\
d\te_i^\al \ved d\teb^{j\ald} & = &d\teb^{j\ald} \ved d\te_i^\al .
\eeqaa

Obviously, to each superfield one can associate an integral
form and write the action (\ref{acg}) in terms of these forms. This would  not give
a new  insight. However, we can collect  differential forms
possessing different degrees in a ``master" superfield as

\newpage

\beqa
\Omo & = &  c_1 (u^i_\al +v^i_\al ) d\te_i^\al 
+ic_2 (2A_{\al\ald}d\te_i^\al \ved d\teb^{i\ald} 
+3\Phi^{ij}\ep_{\al\bet}d\te_i^\al \ved d\te_j^\bet  ) \nonumber \\
&&-2ic_3 (
\La_{i\al} \ep_{\ald\betd}d\te_j^\al\ved d\teb^{i\ald} \ved d\teb^{j\betd}
+4\Lab^i_\ald \ep_{\al\bet} d\teb^{j\ald} \ved d\te_i^\al \ved d\te_j^\bet ) \nonumber \\
&&
+2c_4(2F_{\al\bet} \ep_{\ald\betd}
d\te^{\al}_i\ved d\te_j^\bet\ved d\teb^{i\ald} \ved d\teb^{j\betd}
+8F_{\ald\betd}\ep_{\al\bet}
d\teb^{i\ald} \ved d\teb^{j\betd} \ved d\te_i^\al \ved d\te_j^\bet  \nonumber \\
&&
\qquad\quad+3[ \Phi_{ik},\Phi^{kj}]\ep_{\al\bet}\ep_{\ald\betd} 
d\te_j^\al \ved d\te_n^\bet\ved d\teb^{i\ald} \ved d\teb^{n\betd} ). \label{OO}
\eeqa
Construction of this master superfield is twofold: Firstly, each component is chosen to possess mass dimension equal to its form degree, e.g. the first component is a one form and it has mass dimension one. Secondly, once the first order components
are chosen as one form,
the second order components are related to the first ones
up to the constant $c_2,$ by
the recursion relations 
(\ref{reco}) replacing in the right hand side explicit $\te,\teb$ with the differentials $d\te , d \teb .$ 
The third order ones are obtained from the second order components utilizing the
recursion relations
(\ref{reca}), (\ref{recp}),  up to the constant $c_3 .$ Similarly, the fourth order terms are derived 
from the recursion relation (\ref{recl}) of the third order components, up to the constant
$c_4 .$

To write an action we also need the hermitian conjugate 
of $\Omo :$
\beqa
\Omo^\dagger & = &  -c_1  ({\bar{u}}_{i\ald} +{\bar{v}}_{i\ald})d\teb^{i\ald}
+ic_2 (2A_{\al\ald}d\te_i^\al \ved d\teb^{i\ald} 
-3\Phi_{ij}\ep_{\ald\betd}d\teb^{i\ald} \ved d\teb^{j\betd} )\\
&&
-2ic_3 (
4\La_{i\al}\ep_{\ald\betd} d\te_j^\al\ved d\teb^{i\ald} \ved d\teb^{j\betd}
+\Lab^i_\ald \ep_{\al\bet} d\teb^{j\ald} \ved d\te_i^\al \ved d\te_j^\bet) \nonumber \\
&&-2c_4(8F_{\al\bet} \ep_{\ald\betd}
d\te^{\al}_i \ved d\te_j^\bet\ved d\teb^{i\ald} \ved d\teb^{j\betd}
+2F_{\ald\betd}\ep_{\al\bet}
d\teb^{i\ald} \ved d\teb^{j\betd} \ved d\te_i^\al \ved d\te_j^\bet  \nonumber \\
&&
\qquad \quad+3[ \Phi_{ik},\Phi^{kj}]\ep_{\al\bet}\ep_{\ald\betd} 
d\te_j^\al \ved d\te_n^\bet\ved d\teb^{i\ald} \ved d\teb^{n\betd} ).
\eeqa
Let us introduce the operator
\[
d=i\del_{\al\ald}d\te_i^\al \ved d\teb^{i\ald}
\]
which corresponds to  derivatives  $\del /\del x^\mu.$
In terms of the constants $m_1,m_2,m_3,$ we propose the action, suppressing superspace integrals and trace over the gauge group, 
\be
\label{acf}
I= m_1\Omo^\dagger \ved d \ved\Omo + m_2 \Omo^\dagger \ved \Omo
+ m_3 \left( \Omo \ved \Omo^\dagger \ved \Omo +
\Omo^\dagger \ved \Omo \ved \Omo^\dagger \right)
\ee
and the $SU(4)$ invariant 4--form 
\be
d\te_i^\al \ved d\te_j^\bet \ved d\teb^{k\ald} \ved d\teb^{l\betd}
= \ep^{\al\bet}\ep^{\ald\betd}(\kd^k_i\kd^l_j -\kd^k_j\kd^l_i)
d\te_m^\ga d\te_{n\ga} d\teb^m_\gad d\teb^{n\gad} .
\ee
All other powers of the superfields $\Omo , \Omo^\dagger$ 
give  vanishing contributions due to the choice of the measure (\ref{mes0})--(\ref{mes}).
Comparing the coefficients of (\ref{acf}) with (\ref{acg})
one can show that they are related as
\beqa
k_1=3m_1c_1^2,& k_2=-12m_3c_1^2c_2, & k_3=-48m_2c_1c_3, \nonumber \\
k_4=-48m_2c_2^2,& k_5= -6m_3c_1^2c_2, & k_6=72m_2c_2^2. \label{ceq}
\eeqa
$c_4$ does not play any role.
Note that in this case the condition (\ref{cso}), namely
\[
k_6 = -\frac{3k_4}{2} 
\]
is dictated 	spontaneously.
Replacing $k_1,\cdots , k_6$ in (\ref{acg}) with the values given 
in (\ref{ceq}), one obtains the equations which $c_1,\cdots,m_3$
coefficients should satisfy, so that (\ref{acf}) reproduces the action (\ref{n4c}). There 
exist
several solutions to these equations.  By setting
\[
c_1=c_2=1
\]
one can show that there exists a solution such that
\beqa
c_3 & = & 4+\frac{12}{5} b \left(4+b^2\right),\\
m_1 & = & -\frac{8}{3} (104+b (282+b (16+63 b))),\\
m_2 & = & \frac{1}{16} \left(21+4 b^2\right),\\
m_3 & = &\frac{1}{120} (2815+4 b (1974+b (115+441 b))),
\eeqa
where $b$ and $l$ are given with (\ref{scale}) as before.

\section{Discussions}

We presented a superfield formulation of
N=4 SYM theory in 4 dimensions. 
Superfields which we deal with do not possess auxiliary fields,
in contrast to the standard superfields which one engages to formulate  off--shell supersymmetric theories. 
Thus, techniques to carry out calculations like taking
variations or performing path integrals of their functionals with respect to these superfields are obscure at the moment. 
Hence, we also do not know how to imply supersymmetry invariance of the action (\ref{acg}) at the level of superfields. In spite of all these facts, being able to introduce integral forms to 
write the action (\ref{acg}) in terms of the master superfield $\Omo$ (\ref{acf}) is very 
promising.
Getting a better knowledge of the geometrical aspects of the master field (\ref{OO})  can 
shed some light on the use of our formalism.
One of the
tools to deepen the understanding how to operate with these superfields is to
study the analogous formulations of the N=1 SYM in 4  and 10 dimensions. 
Although, the former is available the latter case is still missing.

Possessing a superfield formulation of N=4 SYM, even though without auxiliary fields, should also be helpful to study deformations of it in terms of Moyal brackets: In spite of the fact that deformed equations of motion of N=4 SYM were worked out \cite{sw}
an underlying action is still missing.

\begin{center}
Acknowledgments
\end{center}

\noindent
We thank M. Hinczewski for fruitful discussions.
Preliminary results of this work were announced 
during the meeting in honor of \mbox{\.{I}.H. Duru} on the occasion of his 60th birthday
at \.{I}YTE, \.{I}zmir, Turkey.

\appendix

\section{Higher components of superfields}

Here we list  components of the superfields which 
are not given in Section 3. Because of the choice of  measure (\ref{mes0})--(\ref{mes})
some of the terms of components evidently 
give vanishing contribution to the action (\ref{acg}).
Below, $``\cdots"$ indicates these terms
which are not needed for our
calculation.

From (\ref{a1}) 
by making use of the recursion relations (\ref{reca})--(\ref{recl}) or  
performing the supersymmetry transformations (\ref{dt}), the  components of
the superfield $A_{\al\ald}$ can be obtained as 
\beqa
A^{(2)}_{\al\ald} & = & -i\left( \te_{i\al}\teb^{i\betd}f_{\ald\betd}+\teb^i_\ald\te_i^\bet f_{\al\bet} 
-b^2 \te_{i\al}\teb^j_\ald [ \phi^{ik},\phi_{kj} ]  \right) +\cdots , \\
A_{\al\ald}^{(3)} & = & 
-\fr{il}{6} \teb^i_\ald \te_i^\bet 
\Big( \te_{k\bet} \rD_{\al\betd} \lab^{k\betd}
+\te_{k\al} \rD_{\bet\betd} \lab^{k\betd} 
+\teb^{k\betd}\rD_{\al\betd}\la_{k\bet}  
+\teb^{k\betd}\rD_{\bet\betd}\la_{k\al} 
+b\te_{j\al}[\la_{k\bet},\phi^{kj}] \Big) \nonumber \\
&& -\fr{il}{6} \te_{i\al} \teb^{i\betd} 
\Big(\teb^k_\ald \rD_{\bet\betd} \la^\bet_k
+\teb^k_\betd \rD_{\bet\ald} \la_k^\bet
+\te_k^\bet \rD_{\bet \betd}\la_\ald^k 
+\te_k^\bet \rD_{\bet \ald} \lab^k_\betd 
-b\teb^j_\ald[\lab^k_\betd,\phi_{kj}] \Big) 
\Big)  +\cdots ,\\
A_{\al\ald}^{(4)} & = & 
-\fr{i}{24} \te_{i\al}\teb^{i\betd}\Big(  
l^2\te_{k\bet} \left(\teb^j_\ald \{\lab^k_\betd,\la_j^\bet\}+\teb^j_\betd \{\lab^k_\ald,\la_j^\bet\}\right) 
-l^2\te_{j}^\bet \left(\teb^k_\betd \{\la_{k\bet},\lab^j_\ald\}+\teb^k_\ald \{\la_{k\bet},\lab^j_\betd\}\right) \nonumber \\
&&\qquad\quad +2i\te_{j\ga} \left(\teb^j_\ald \rD_{\bet\betd}f^{\bet\ga}
+\teb^j_\betd \rD_{\bet\ald}f^{\bet\ga}\right) 
+2i\te_j^\bet \teb^{j\gad} \left( \rD_{\bet\betd}f_{\ald\gad}
+\rD_{\bet\ald}f_{\betd\gad}\right) \nonumber \\
&&\qquad\quad +\teb^j_{\ald}\left(-4ib^2\te_m^\bet [\rD_{\bet\betd}\phi^{km},\phi_{kj}]
+2l^2\te_k^\bet \{\lab^k_\betd,\la_{j\bet}\}
-2l^2\te_j^\bet \{\lab^k_\betd,\la_{k\bet} \}\right) \Big) \nonumber \\
&&  -\fr{i}{24} \teb^i_{\ald}\te^{\bet}_i\Big( 
l^2\teb^k_{\betd} \left(\te_{j\bet} \{\la_{k\al},\lab^{j\betd}\}+\te_{j\al} \{\lab^j_\betd,\la_{k\bet}\}\right) 
+l^2\teb^{j}_\betd \left( \te_{k\al} \{\la_{j\bet},\lab^{k\betd}\}+\te_{k\bet} \{\la_{j\al},\lab^k_\betd\}\right) \nonumber \\ 
&&\qquad\quad +2i\teb^j_{\gad} \left( \te_{j\bet} \rD_{\al\betd}f^{\betd\gad}+\te_{j\al} \rD_{\bet \betd}f^{\betd\gad} \right) +2i\teb^{j\betd} \te_j^\ga \left( \rD_{\bet\betd}f_{\al\ga}+\rD_{\al\betd}f_{\bet\ga}\right) \nonumber \\
&&\qquad\quad +\te_{j\al}\left(-4ib^2\teb^{m\betd} [\rD_{\bet\betd}\phi_{km},\phi^{kj}]
+2l^2\teb^{k\betd} \{ \la_{k\bet},\lab^j_\betd \}-2l^2\teb^{j\betd} \{ \lab^k_\betd,\la_{k\bet} \right) \Big) +\cdots .
\eeqa

To derive  components of the superfield $\La_{i\al}$ 
one departs from (\ref{l1}) and 
uses the recursion relations (\ref{reca})--(\ref{recl}) or  
performs the supersymmetry transformations (\ref{dt}):
\beqa
\La^{(2)}_{i\al} & = & \fr{il}{2} \te_i^\bet \Big( \te_{k\bet} \rD_{\al\ald} \lab^{k\ald}
+\te_{k\al} \rD_{\bet\ald} \lab^{k\ald} 
+3\teb^{k\ald}\rD_{\al\ald}\la_{k\bet}  \nonumber \\
& & +\teb^{k\ald}\rD_{\bet\ald}\la_{k\al} 
+b\te_{j\al}[\la_{k\bet},\phi^{kj}] \Big) \nonumber \\
&&+\fr{ilb}{2}\te_{j\al} \left(
\ep_{iklm} \teb^{l\ald}[\lab^m_{\ald} \phi^{kj}]
-\teb^{k\ald}[\lab^j_{\ald},\phi_{ik}]
+\teb^{j\ald}[\lab^k_{\ald},\phi_{ik}]
\right) \nonumber \\
&& +il\teb^{j\ald} \left(
b\te_{k\al}[\lab^k_{\ald}, \phi_{ij}]
+ b\teb^k_\ald [\la_{k\al},\phi_{ij} ]
+\te_j^\bet \rD_{\al\ald} \la_{i\bet} +\cdots .
\right) +\cdots , \\
\La^{(3)}_{i\al} & = &\fr{i}{6}\Big(
\te_i^\bet\teb_\ald^m 
( 
2l^2\te_{k\bet}\{\la_{m\al},\lab^{k\ald}\} +8l^2\te_{k\al}\{\la_{m\bet},\lab^{k\ald} \} 
-2l^2\te_{m\al} \{\la_{k\bet},\lab^{k\ald} \} \nonumber \\
&&
+6l^2\teb^{k\ald}\{\la_{m\al},\lab_{k\bet} \}
+2i\te_{m\bet}\rD_{\al\betd} f^{\ald\betd} +2i\te_{m\al}\rD_{\bet\betd} f^{\ald\betd}) \nonumber \\
&& +2i\te_i^\bet\teb^{m \ald}
( \te_m^\ga \rD_{\bet \ald}f_{\al\ga}+2b^2 \te_{k\bet}\rD_{\al\ald}[\phi^{kj},\phi_{jm}]
+b^2\te_{k\al}\rD_{\bet\ald}[\phi^{kj},\phi_{jm}]+b^2\te_{j\al}[\rD_{\bet\ald}\phi_{km},\phi^{kj}]) \nonumber \\
&&-l^2\teb^{j\ald}(5\te_j^\bet\te_{k\al}+\te_{j\al}\te_k^\bet )\{\la_{i\bet},\lab^k_\ald\} 
\nonumber \\
&& -2l^2\ep_{iklm}\teb^{j\ald}\te_{j\al}\teb^{l\betd}\{ \lab^m_\betd ,\lab^k_\ald\}
-4l^2\teb^{j\ald}\te_j^\bet\teb^{k\ald} \{ \la_{k\al} ,\la_{i\bet} \}\nonumber \\
&&
+2ib^2\te_{j\al}\te_n^\ga ( 
\ep_{iklm}\teb^{l\ald} [\rD_{\ga \ald}\phi^{mn},\phi^{kj}] 
+\teb^{j\ald} [\rD_{\ga \ald}\phi^{nk},\phi_{ik}])
\nonumber \\
&&
-2ib\teb^{j\ald}\te_j^\bet (b\te_{m\bet} \rD_{\al\ald}[\phi_{ik},\phi^{km}]
+2\teb^{k\betd}\rD_{\al\ald}\rD_{\bet\betd} \phi_{ik} )
\nonumber \\
&&
+4ib \teb^{j\ald}\teb^k_\ald\te_k^\bet[f_{\al\bet},\phi_{ij}]
\nonumber \\
&&
+ib^3\te_{j\al}\teb^m_\ald (\ep_{ikln}\teb^{l\ald}[[\phi^{np},\phi_{pm}],\phi^{kj}]
+\teb^{k\ald}[[\phi^{nj},\phi_{nm}],\phi_{ik}] )
\nonumber \\
&&
+ib^3\teb^{j\ald}\teb^m_\ald (\te_{j\al} [[\phi^{nk},\phi_{nm}],\phi_{ik}]
+4\te_{k\al}[[\phi^{kn},\phi_{nm}],\phi_{ij}])
\Big) +\cdots .
\eeqa

Similarly  components of the superfield $\Phi^{ij}$ are
calculated from (\ref{p1}) in terms of
the recursion relations (\ref{reca})--(\ref{recl}) or  
the   supersymmetry transformations (\ref{dt})  as
\beqa
\Phi^{(2)ij} & = & -\fr{i}{2}\Big(
2b\teb^i_\ald \te_k^\al \rD_{\al\ald} \phi^{jk}
-b\ep^{ijkl}\teb^{n\ald} \te^\al_k \rD_{\al\ald} \phi_{ln}
-b^2\teb^i_\ald \teb^{m\ald}[\phi^{jk},\phi_{km}] \nonumber \\
& &
+\fr{b^2}{2}\ep^{ijkl}\te_k^\al \te_{m\al}[\phi_{ln},\phi^{nm}]
\Big)+\fr{i}{2}\Big(i\longleftrightarrow j\Big) +\cdots ,\\
\Phi^{(3)ij} & = & \fr{i}{6}\Big(
2l\teb^{i\ald}\teb^{k\betd} \te_k^\al \rD_{\al\ald} \lab^j_\betd
-l\ep^{ijkl}\teb^{m\ald}\te_k^\al \te_m^\bet \rD_{\al\ald} \la_{l\bet}
+lb\ep^{jkln} \teb^{i\ald} \teb^{m}_{\ald} \te^{\al}_{l} 
[\la_{n\al},\phi_{km}] \nonumber \\
& &
+lb \teb^{i\ald} \teb^{m}_{\ald} \te^{\al}_{m}
[\la_{k\al},\phi^{jk}]
-3lb \teb^{i\ald} \teb^{m}_{\ald} \te^{\al}_{k}
[\la_{m\al},\phi^{jk}]
+lb \ep^{ijkl} \teb^{m\ald} \teb^{n}_{\ald} \te^{\al}_{k}
[\la_{n\al},\phi_{lm}]  \nonumber \\
& &
+2lb \te_k^\al \te_{m\al} \teb^{i\ald}
[\lab_{\ald}^m,\phi^{jk}]
+\fr{lb}{2} \ep^{ijkl}\ep_{lnpr} \te^{\al}_{k} \te_{m\al} \teb^{p\betd}
[\lab_{\betd}^r,\phi^{nm}]
-\fr{3lb}{2} \ep^{ijkl}\te^{\al}_{k} \te_{m\al} \teb^{n\betd}
[\lab_{\betd}^m,\phi_{ln}] \nonumber \\
& &
+\fr{lb}{2}\ep^{ijkl} \te^{\al}_{k}  \te_{m\al} \teb^{m\betd}
[\lab_{\betd}^n,\phi_{ln}]
\Big)-\fr{i}{6}\Big(i\longleftrightarrow j\Big) +\cdots ,\\
\Phi^{(4)ij} & = & \fr{1}{24}\Big(
\ep^{jkln}\teb^{i\ald}\teb^m_\ald\te_l^\al\left(b^3\te_{p\al}
[[\phi_{nr},\phi^{rp}],\phi_{km}] 
+2il^2 \te_m^\bet \{\la_{n\al},\la_{k\bet}\}\right)
-4il^2\teb^{i\ald}\teb^{k\betd}\te_k^\al\te_{m\al}
[\lab^m_\ald ,\lab^j_\betd ]
\nonumber \\
& &
-b^3\teb^{i\ald}\teb^m_\ald \te_{n\al}\left(
5\te_k^\al  [\phi^{jk},[\phi_{mr},\phi^{rn}]]
-\te_m^\al  [\phi^{jk},[\phi_{kr},\phi^{rn}]] \right) \nonumber \\
& &
+\fr{1}{2}\ep^{ijkl} \te_k^\al \te_{m\al}\big(b^3 \ep_{lnpr} 
 \teb^{p\ald}\teb^q_\ald [[\phi^{rz},\phi_{zq}],\phi^{nm}]
+2il^2\ep_{lnpr}\teb^{m\ald}\teb^{p\betd} \{\lab^n_\ald ,\lab^r_\betd\}
\nonumber \\
& &
+5b^3\teb^{n\ald}\teb^p_\ald [\phi_{ln},[\phi^{mr},\phi_{rp}]]
-b^3\teb^{m\ald}\teb^p_\ald [\phi_{ln},[p\phi^{nr},\phi_{rp}]]
\big)+2il^2 \ep^{ijkl} \te_k^\al\te_m^\bet \teb^{m\ald} \teb^n_\ald
\{\la_{n\al},\la_{l\bet}\}
\Big)\nonumber \\
& &
-4 b \teb^{i\ald}\teb^{k\betd}\te_k^\al\te_{l}^{\bet}\rD_{\al\ald}\rD_{\bet\betd}\phi^{jl} 
-2\epsilon^{ijkl} \teb^{m\ald}\teb^{n\betd}\te_k^\al\te_{m}^{\bet}\rD_{\al\ald}\rD_{\bet\betd}\phi_{ln}\Big)\nonumber\\
&&-\fr{1}{24}\Big(i\longleftrightarrow j\Big) +\cdots .
\eeqa

To find the third and fourth order components in $\te,\teb$ of the spinor superconnections
$\omo_\al^i\equiv v_i^\al+u_\al^i$ one takes  (\ref{o1})--(\ref{o2})
and operates with the recursion relations (\ref{reca})--(\ref{recl}):
\beqa
v^{(3)i}_\al &  = & \fr{1}{2}\teb^{i\ald}\left( \teb^j_\ald \te_j^\bet f_{\al\bet}
+b^2 \teb^m_\ald \te_{j\al} [\phi^{jk},\phi_{km}] \right) +\cdots , \\
 u^{(3)i}_\al & = & -\fr{b^2}{4}\te_{j\al}\teb^m_{\ald}\left( \teb^{i\ald}
[\phi^{jk},\phi_{km}]-\teb^{j\ald} [\phi^{ik},\phi_{km}] \right)
-2b \teb^{j\betd}\te_{j\al}\te_k^\bet \rD_{\bet \betd} \phi^{ik} +\cdots , \\
v^{(4)i}_\al &  = & -\fr{l}{15}\teb^{i\gad}\te_j^\bet \Big(
\teb^{k\betd}\te_{k\al}
(\rD_{\bet\betd}\lab^j_\gad + \rD_{\bet\gad}\lab^j_\betd )
+\teb^j_\gad \te_{k\bet} \rD_{\al\betd}\lab^{k\betd} \nonumber \\
& &  
+\teb^j_\gad \te_{k\al} \rD_{\bet\betd}\lab^{k\betd}
+b\te_\gad^j\te_{m\al}[\phi^{mk},\la_{k\bet}]
\Big)+\cdots ,    \\
 u^{(4)i}_\al  & = &  -\fr{l}{15} \te_{j\al}\teb^m_\ald \Big(
 b\te^{i\ald} [\phi^{jk},\la_{k\bet}] -\te_m^\bet ( b \teb^{j\ald} [\phi^{ik},\la_{k\bet}]
-2\teb^{i\gad} \rD_{\bet\gad} \lab^{j\ald}+2\teb^{j\gad}\rD_{\bet\gad}\lab^{i\ald} ) \nonumber \\
&& 
-3b\te_k^\bet \teb^{j\ald} [\phi^{ik},\la_{m\bet}] 
-b\ep^{ikln}\te_l^\bet \teb^{j\ald} [ \la_{n\bet},\phi_{km} ] \Big)
+\cdots .
\eeqa

The higher components in $\te ,\teb$ which are not listed here do not play any role in our calculations.

\end{document}